\documentclass[%
 reprint,
 amsmath,amssymb,
 aps, 
 prd,
 longbibliography,
]{revtex4-1}
 
\usepackage{mathtools}
\usepackage{graphicx}% Include figure files
\usepackage{dcolumn}% Align table columns on decimal point
\usepackage{bm}% bold math
\begin{document}

\preprint{APS/123-QED}

\title{Differences in binarity between gluon and graviton scattering amplitudes}% Force line breaks with \\

\author{Abhilash \hspace{-0.3cm} Mathews}
%  \altaffiliation[Also at ]{Physics Department, XYZ University.}%Lines break automatically or can be forced with \\
 \email{mathewsa@mit.edu}
\affiliation{%
\vspace{0.1cm}
Plasma Science and Fusion Center, Massachusetts Institute of Technology, 77 Massachusetts Avenue, Cambridge, Massachusetts 02139, United States
}%

% \date{November 24, 2019}
\date{\today}% It is always \today, today,

%A simple, quick, dense review of remarkable results derived through on-shell amplitudes for novices based upon references contained
\begin{abstract} 
\noindent Both 3- and 4-point scattering amplitudes for spin-1 massless particles (gluons) and spin-2 massless particles (gravitons) are reviewed through self-contained step-by-step derivation. Gluon and graviton interactions are computed from on-shell diagrams by starting from complex momenta and introducing spinor-helicity formalism. By Fourier transforming spinor variables in momentum space, binarity is revealed in pure Yang-Mills scattering amplitudes while absent to a certain extent for gravity indicating fundamental differences in the probability spaces spanned by the two interactions.

% 'amplitudes that are directly constructed through on-shell methods are automatically gauge-invariant.
% This simple fact dramatically increases both (a) the computational simplicity of calculations
% of scattering amplitudes, and (b) the physical transparency of the final results.'
% \begin{description}
% Secondary publications and information retrieval purposes.
% You may use the \texttt{description} environment to structure your abstract;
% use the optional argument of the \verb+\item+ command to give the category of each item. 
% \end{description}
\end{abstract}

\maketitle

%\tableofcontents

\section{\label{sec:level1} Spinor-helicity formalism:\protect\\ Introducing particles and momenta}

\noindent A particle state is uniquely defined by invariance under transformations of the Poincar\'e group \cite{Weinberg:1995mt,arkanihamed2017scattering} (e.g. translations, spatial rotations, boosts) and can be identified by its momentum, $p$, and intrinsic quantum numbers (e.g. spin, parity, charge), where helicity, $h$, is the particle's spin projected along its direction of motion, i.e. momentum vector \cite{Benincasa2007xk}. In natural units, the complex momentum of a massless particle can be described by the following four-vector and conditions without loss of generality \cite{marzolla2017fourdimensional}:
\begin{equation}
    p_\mu = \begin{pmatrix} p_0 \\ p_1 \\ p_2 \\ p_3 \end{pmatrix} = \begin{pmatrix}E \\ 0 \\ 0 \\ E\end{pmatrix} \ ; \ p^2 = p^\mu p_\mu = m^2 = 0
\end{equation}
This four-vector can be directly mapped into a $2 \times 2$ matrix representation via the Pauli spin matrices \cite{cheung2017tasi}
\begin{equation}
    p_{\alpha \dot{\alpha}} = p_\mu \sigma^\mu_{\alpha \dot{\alpha}} = \begin{pmatrix} p_0 + p_3 & p_1 - ip_2 \\ p_1 + ip_2 & p_0 - p_3 \end{pmatrix}
\end{equation}
where
\begin{equation*}
  \sigma^0 =
    \begin{pmatrix}
      1&0\\
      0&1
    \end{pmatrix}, \\
  \sigma^1 =
    \begin{pmatrix}
      0&1\\
      1&0
    \end{pmatrix}, 
    \\
  \sigma^2 =
    \begin{pmatrix}
      0&-i\\
      i&0
    \end{pmatrix}, \\
  \sigma^3 =
    \begin{pmatrix}
      1&0\\
      0&-1
    \end{pmatrix} \
\end{equation*}
and the Lorentz invariant quantity 
\begin{equation} \label{invar-lorentz}
p^2 = p^\mu p_\mu = \det(p) = 0
\end{equation}
is made manifest by the matrix's construction. Given a particle state, $\vert p , h \rangle$, any general Lorentz transformation from momentum $p$ to $p'$ is denoted as $L(p';p) \equiv \Lambda$. Assuming the existence of a unitary representation, $U(L(p';p))$, particle states evolve as
\begin{equation}
\vert p , h \rangle = U(L(p;k)) \vert k , h \rangle
\end{equation}
This statement can be equivalently written as
\begin{multline}
U(\Lambda) \vert p , h \rangle = U(\Lambda) U(L(p;k)) \vert k , h \rangle \\
= U(L(\Lambda p;k)) U(\underbrace {L^{-1}(\Lambda p;k) \Lambda L(p;k)}_{\mathclap{W(\Lambda,p,k)}}) \vert k , h \rangle 
\end{multline}
where the transformations which leave momentum invariant, $W$, are known as `little group' transformations \cite{wigner_little,bargman_pt2}. These are critically important when formulating scattering amplitudes via the spinor-helicity formalism because of the weight (phase for real momenta) that is added even if momentum is invariant. This will now be outlined by deconstructing $p_{\alpha \dot{\alpha}}$ into its spinor representation \cite{marzolla2017fourdimensional,cheung2017tasi}:
\begin{equation}
    p_{\alpha \dot{\alpha}} = \lambda_\alpha \otimes \tilde{\lambda}_{\dot{\alpha}} = \begin{pmatrix} \lambda_1\tilde{\lambda}_1 & \lambda_1\tilde{\lambda}_2 \\ \lambda_2\tilde{\lambda}_1 & \lambda_2\tilde{\lambda}_2 \end{pmatrix} \equiv \lambda_\alpha \tilde{\lambda}_{\dot{\alpha}}
\end{equation}
\begin{equation}
    \text{e.g. } \lambda_\alpha = \sqrt{2E} \begin{pmatrix} 1 \\ 0 \end{pmatrix} \ ; \ \tilde{\lambda}_{\dot{\alpha}} = \sqrt{2E} \begin{pmatrix} 1 \\ 0 \end{pmatrix}
\end{equation}

\noindent The following notation for particles $i$ and $j$ is defined by
\begin{equation}
    \langle i j \rangle \equiv \lambda^{(i)}_\alpha \lambda^{(j)}_\beta \varepsilon^{\alpha \beta} \ ; \
    [ i j ] \equiv \tilde{\lambda}^{(i)}_{\dot{\alpha}} \tilde{\lambda}^{(j)}_{\dot{\beta}} \varepsilon^{\dot{\alpha} \dot{\beta}}
\end{equation}
where $\varepsilon^{ij}$ ($\varepsilon_{ij}$) is the two-dimensional contravariant (covariant) Levi-Civita tensor:
\begin{equation}
    \varepsilon^{ij} = -\varepsilon_{ij} =
        \begin{pmatrix}
      0&1\\
      -1&0
    \end{pmatrix}
%   \begin{cases}
%   -1       & \text{if } (i, j) = (1, 2) \\
%   +1       & \text{if } (i, j) = (2, 1) \\
%     \;\;\,0 & \text{if } i = j
%   \end{cases}
\end{equation}
and by its antisymmetry determines the below relations
\begin{equation*}
    \langle i j \rangle = -\langle j i \rangle, \langle i i \rangle = 0, [ i j ] = -[ j i ], [ j j ] = 0
\end{equation*}
Consequently, the Lorentz invariant quantity for massless particles $i$ and $j$ can be expressed as the following due to momentum conservation and the on-shell condition
% In physics, particularly in quantum field theory, configurations of a physical system that satisfy classical equations of motion are called on shell, and those that do not are called off shell.
% In quantum field theory, virtual particles are termed off shell because they do not satisfy the energy–momentum relation; real exchange particles do satisfy this relation and are termed on shell (mass shell).[1] In classical mechanics for instance, in the action formulation, extremal solutions to the variational principle are on shell and the Euler–Lagrange equations give the on-shell equations. Noether's theorem regarding differentiable symmetries of physical action and conservation laws is another on-shell theorem.
\begin{equation}
\label{on-shell}
     p^{(i)} \cdot p^{(j)} = \langle i j \rangle [ i j ] = 0
\end{equation}
which implies either $\langle i j \rangle = 0$ (i.e. $\lambda^{(i)}_{\alpha} || \lambda^{(j)}_{\alpha}$) or $[ i j ] = 0$ (i.e. $\tilde{\lambda}^{(i)}_{\dot{\alpha}} || \tilde{\lambda}^{(j)}_{\dot{\alpha}}$) for non-trivial $p^{(i)}$. To reiterate, this is explicitly momentum conservation for massless particles:
\begin{equation}
     \sum\limits_{i} p^{(i)} = \sum\limits_{i} \lambda^{(i)}_\alpha \tilde{\lambda}^{(i)}_{\dot{\alpha}} = 0 
\end{equation}
Complex momenta are not uniquely defined through this representation because a re-scaling of spinors via
\begin{equation*}
\lambda_\alpha \rightarrow t \lambda_\alpha \ \text{and} \ \tilde{\lambda}_{\dot{\alpha}} \rightarrow t^{-1} \tilde{\lambda}_{\dot{\alpha}}
\end{equation*}
leaves momentum invariant. This corresponds exactly to the little group action, where general Lorentz transformations in spinor-helicity variables become
\begin{equation}
p'_{\alpha \dot{\alpha}} = L(p';p)_\alpha^\beta \tilde{L}(p';p)_{\dot{\alpha}}^{\dot{\beta}} \lambda_\beta \tilde{\lambda}_{\dot{\beta}}
\end{equation}
Note that this framework can be extended to massive particles (and naturally derives various limiting forms, e.g. Dirac equation), too \cite{arkanihamed2017scattering}. With prescient formalism now defined for massless particles $i$ and $j$, scattering amplitudes void of virtual particles are ready to be computed.
\section{\label{sec:level2} Three-point scattering \\ in momentum and twistor space}
\begin{figure}
\includegraphics{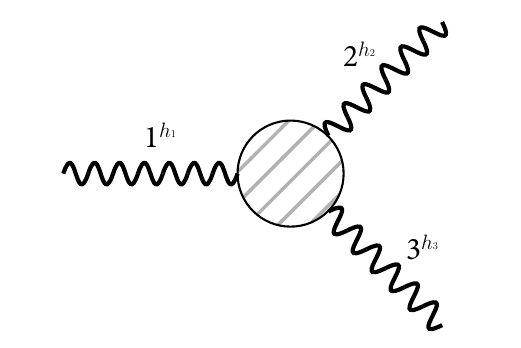}% Here is how to import EPS art
\caption{\label{fig:feyn_diag3} Example of a three particle scattering interaction.}
\end{figure}
% https://stats.stackexchange.com/questions/9617/what-is-the-purpose-of-characteristic-functions
\noindent The S-matrix determines the probability amplitude of incoming particles in an interaction transforming into outgoing particles. The elements composing the S-matrix, $S_{fi}$, are known as scattering amplitudes and are fundamentally used to construct experimental observables (e.g. differential cross-sections of reactions) by assessing the probability obtained by taking their modulus squared, i.e. $\lvert S_{fi} \rvert^2 $. Therefore, ascertaining the structure of scattering amplitudes assists predictive capability and understanding the nature of where allowed particle interactions exist. For example, consider a 3-point interaction as depicted by Figure 1. Scattering amplitudes are generically functions of particles' momenta and helicities \cite{Benincasa2007xk}
\begin{equation}
\hspace{-0.12cm}
    S_{fi} \rightarrow A_{123} = f (\{\lambda^{(i)}_\alpha, \tilde{\lambda}^{(i)}_{\dot{\alpha}}, h_i\})
\end{equation}
An initial ansatz is that scattering amplitudes transform under the Poincar\'e group just as individual particles. Therefore, under the helicity operator (i.e. a little group action), the scattering amplitude transforms as \cite{marzolla2017fourdimensional}
\begin{equation}
    (\lambda^{(i)}_{\alpha} \frac{\partial}{\partial \lambda^{(i)}_{\alpha}} - \tilde{\lambda}^{(i)}_{\dot{\alpha}} \frac{\partial}{\partial \tilde{\lambda}^{(i)}_{\dot{\alpha}}}) A_{123} = -2h_i A_{123}
\end{equation}
where the negative sign is introduced by convention. Recall that momentum conservation requires either
\begin{equation}
    \langle 12 \rangle = \langle 23 \rangle = \langle 31 \rangle = 0 \text{ or } [12] = [23] = [31] = 0,
\end{equation}
and full scattering amplitudes are constrained to be functions of purely on-shell quantities, therefore \cite{McGady_2014}:
\begin{equation}
    A_{123} = A_{123}^{(\lambda)} (\langle 12 \rangle , \langle 23 \rangle , \langle 31 \rangle) + A_{123}^{(\tilde{\lambda})} ([12],[23],[31])
\end{equation}
Combining the above ansatzes and conditions prescribes the following scattering amplitude as a solution \cite{cheung2017tasi}:
\begin{equation*}
    A_{123} = 
  \begin{cases}
  g_{123}^- \langle 12 \rangle^{h_3 - h_1 - h_2} \langle 23 \rangle^{h_1 - h_2 - h_3} \langle 31 \rangle^{h_2 - h_1 - h_3}      \\ \text{ \hspace*{140pt} if } \sum
  \limits_{i} h_i < 0 \\
  g_{123}^+ [12]^{h_1 + h_2 - h_3} [23]^{h_2 + h_3 - h_1} [31]^{h_3 + h_1 - h_2}       \\ \text{ \hspace*{140pt} if } \sum\limits_{i} h_i > 0
  \end{cases}
\end{equation*}
where $g^\pm_{123}$ is the associated coupling constant, and the necessary condition $\delta^4 (\sum\limits_{i} \lambda^{(i)}_\alpha \tilde{\lambda}^{(i)}_{\dot{\alpha}})$ is implicitly demanded for all amplitudes throughout the paper to be on-shell. From hereafter, all scattering amplitudes and transformations will have constants dropped for brevity. Therefore, the steps presented are exact up to overall constants. 
\subsection{Gluon scattering}
\noindent Establishing the preceding background and formalism now permits analysis of concrete scenarios. Take for example a scattering process involving three gluons of helicities $h_1 = +1$, $h_2 = +1$, and $h_3 = -1$. This would produce the following amplitude in momentum space:
\begin{multline}
    A^{YM}_{++-} = \frac{[12]^{3}}{[23] [31]} \delta^4 (\lambda^{(1)}_\alpha \tilde{\lambda}^{(1)}_{\dot{\alpha}} + \lambda^{(2)}_\alpha \tilde{\lambda}^{(2)}_{\dot{\alpha}} + \lambda^{(3)}_\alpha \tilde{\lambda}^{(3)}_{\dot{\alpha}}) \\
    = \frac{[12]^{3}}{[23] [31]} \int d^4X^{\alpha \dot{\alpha}} e^{iX^{\alpha \dot{\alpha}} (\sum\limits_{i} \lambda^{(i)}_\alpha \tilde{\lambda}^{(i)}_{\dot{\alpha}})}
\end{multline}
Taking its Fourier transform at points in spacetime, $X^{\alpha \dot{\alpha}} = X^\mu \sigma^{\alpha \dot{\alpha}}_\mu$, with respect to ${\lambda}^{(1)}_{\alpha}$, ${\lambda}^{(2)}_{\alpha}$, and $\tilde{\lambda}^{(3)}_{\dot{\alpha}}$ results in the scattering amplitude's representation in twistor components ($\mu^{\dot{\alpha}(i)} = -X^{\alpha \dot{\alpha}} \lambda^{(i)}_\alpha$ and $\tilde{\mu}^{\alpha(i)} = -X^{\alpha \dot{\alpha}} \tilde{\lambda}^{(i)}_{\dot{\alpha}}$ with indices raised or lowered accordingly) \cite{Witten_2004,Mason_2010}:
\begin{multline}
    \hat{A}^{YM}_{++-} = [12]^3 \int d^4X^{\alpha \dot{\alpha}} d^2{\lambda}^{(1)}_{\alpha} d^2{\lambda}^{(2)}_{\alpha} d^2\tilde{\lambda}^{(3)}_{\dot{\alpha}} \{ 
    \\ \frac{1}{[23][31]}e^{i\tilde{\mu}^{{\alpha}(1)} {\lambda}^{(1)}_{\alpha}} e^{i\tilde{\mu}^{\alpha (2)} {\lambda}^{(2)}_{\alpha}}
    e^{i{\mu}^{\dot{\alpha}(3)} \tilde{\lambda}^{(3)}_{\dot{\alpha}}}
    e^{iX^{\alpha \dot{\alpha}} (\sum\limits_{i} \lambda^{(i)}_\alpha \tilde{\lambda}^{(i)}_{\dot{\alpha}})}
    \}
    \\ \implies
    \hat{A}^{YM}_{++-} = [12]^3 \int d^4X^{\alpha \dot{\alpha}} \{ \delta^2(\tilde{\mu}^{\alpha(1)} + X^{\alpha \dot{\alpha}} \tilde{\lambda}^{(1)}_{\dot{\alpha}}) \\ \delta^2(\tilde{\mu}^{\alpha(2)} + X^{\alpha \dot{\alpha}} \tilde{\lambda}^{(2)}_{\dot{\alpha}}) 
    \int \frac{d^2\tilde{\lambda}^{(3)}_{\dot{\alpha}}}{[23][31]}
    e^{i\tilde{\lambda}^{(3)}_{\dot{\alpha}}(\mu^{\dot{\alpha}(3)} + X^{\alpha \dot{\alpha}} {\lambda}^{(3)}_\alpha)} \} 
\end{multline}
Applying the coordinate transformation 
\begin{equation}
    \tilde{\lambda}^{(3)}_{\dot{\alpha}} = a_1 \tilde{\lambda}^{(1)}_{\dot{\alpha}} + a_2 \tilde{\lambda}^{(2)}_{\dot{\alpha}}
\end{equation}
and computing its Jacobian changes variables as \cite{Arkani_Hamed_2010}
\begin{equation}
    d^2\tilde{\lambda}^{(3)}_{\dot{\alpha}} = \bigg| \frac {\partial(\tilde{\lambda}^{(3)}_{\dot{1}}, \tilde{\lambda}^{(3)}_{\dot{2}})}{\partial(a_1,a_2)} \bigg| da_1 da_2 = \text{sgn}([12])[12]da_1 da_2
\end{equation} 
\begin{multline}
\label{a_ym_init}
    \implies \hat{A}^{YM}_{++-} = [12]^3 \int d^4X^{\alpha \dot{\alpha}} d^2{\lambda}^{(1)}_{\alpha} d^2{\lambda}^{(2)}_{\alpha} da_1 da_2
  \{ 
    \\ \frac{\text{sgn}([12])[12]}{[23][31]}e^{i\tilde{\mu}^{\alpha (1)} {\lambda}^{(1)}_{\alpha}} e^{i\tilde{\mu}^{\alpha (2)} {\lambda}^{(2)}_{\alpha}}
    e^{i{\mu}^{\dot{\alpha}(3)} (a_1 \tilde{\lambda}^{(1)}_{\dot{\alpha}} + a_2 \tilde{\lambda}^{(2)}_{\dot{\alpha}})} \\
    e^{iX^{\alpha \dot{\alpha}} (\lambda^{(1)}_\alpha \tilde{\lambda}^{(1)}_{\dot{\alpha}} + \lambda^{(2)}_\alpha \tilde{\lambda}^{(2)}_{\dot{\alpha}})}
    e^{iX^{\alpha \dot{\alpha}}\lambda^{(3)}_\alpha (a_1 \tilde{\lambda}^{(1)}_{\dot{\alpha}} + a_2 \tilde{\lambda}^{(2)}_{\dot{\alpha}})}
    \}
\end{multline}
where, using $[23][31] = a_1 a_2 [12]^2$, (\ref{a_ym_init}) becomes
\begin{multline}
    \hat{A}^{YM}_{++-} = \text{sgn}([12])[12]^2 \int d^4X^{\alpha \dot{\alpha}} \frac{da_1 da_2}{a_1a_2} \{ \\ \delta^2(\tilde{\mu}^{\alpha (1)} + X^{\alpha \dot{\alpha}} \tilde{\lambda}^{(1)}_{\dot{\alpha}})
    e^{ia_1({\mu^{\dot{\alpha}}(3)}\tilde{\lambda}^{(1)}_{\dot{\alpha}} - {\tilde{\mu}^{\alpha}(1)} \lambda^{(3)}_\alpha )} \\
    \delta^2(\tilde{\mu}^{\alpha (2)} + X^{\alpha \dot{\alpha}} \tilde{\lambda}^{(2)}_{\dot{\alpha}})
    e^{ia_2({{\mu}^{\dot{\alpha}(3)}\tilde{\lambda}^{(2)}_{\dot{\alpha}}} - {\tilde{\mu}^{\alpha}(2)} \lambda^{(3)}_\alpha)}
    \}
\end{multline} 
Now using the integral PV$\int^\infty_{-\infty} da \frac {e^{iax}}{ca -b} = \frac{i\pi}{c}e^{ibx/c}\text{sgn}(x)$, where $\text{sgn}(x)$ is the signum function, along with
\begin{multline*}
\hspace{-0.35cm}
    \int d^4X^{\alpha \dot{\alpha}}  \delta^2(\tilde{\mu}^{\alpha (i)} + X^{\alpha \dot{\alpha}} \tilde{\lambda}^{(i)}_{\dot{\alpha}})
    \delta^2(\tilde{\mu}^{\alpha (j)} + X^{\alpha \dot{\alpha}} \tilde{\lambda}^{(j)}_{\dot{\alpha}}) \rightarrow \frac{1}{[ij]^2}
\end{multline*}
while defining the following notation
\begin{equation*} 
    W^{(i)} \equiv \begin{pmatrix} \tilde{\lambda}^{(i)} \\ \tilde{\mu}^{(i)} \end{pmatrix}, 
    Z^{(i)} \equiv \begin{pmatrix} {\mu}^{(i)} \\
    {\lambda}^{(i)}
    \end{pmatrix} 
\end{equation*}
\begin{equation*}
    W^{(i)} \cdot Z^{(j)} \equiv \tilde{\mu}^{\alpha(i)} {\lambda}^{(j)}_{\alpha} - {\mu}^{\dot{\alpha}(j)} \tilde{\lambda}^{(i)}_{\dot{\alpha}}
\end{equation*}
results in the below succinct solution for the full non-perturbative 3-point scattering amplitude of gluons \cite{Arkani_Hamed_2010}:
\begin{equation} 
    \hat{A}^{YM}_{++-} = \text{sgn}([12]) \text{sgn}(W^{(1)} \cdot Z^{(3)})
    \text{sgn}(W^{(2)} \cdot Z^{(3)})
\end{equation}
For completeness, these guided steps are all repeated for the opposite helicity configuration (i.e. $h_1 = -1$, $h_2 = -1$, and $h_3 = +1$) to instead yield the answer as:
\begin{equation} 
    \hat{A}^{YM}_{--+} = \text{sgn}(\langle 1 2 \rangle) \text{sgn}(W^{(3)} \cdot Z^{(1)})
    \text{sgn}(W^{(3)} \cdot Z^{(2)})
\end{equation}
The binarity of the quantum chromodynamic interaction is featured by signum functions outputting either +1 or --1 characteristics which sharply contrasts the non-binary graviton scattering amplitude derived in the next section. 
% In technical terms, gluons are vector gauge bosons that mediate strong interactions of quarks in quantum chromodynamics (QCD). Gluons themselves carry the color charge of the strong interaction. This is unlike the photon, which mediates the electromagnetic interaction but lacks an electric charge. Gluons therefore participate in the strong interaction in addition to mediating it, making QCD significantly harder to analyze than QED (quantum electrodynamics).
\subsection{Graviton scattering}
\noindent From the common starting point of $A_{123}$, suppose a scattering process involves  three  gravitons  of  helicities $h_1 = +2$, $h_2 = +2$, and $h_3 = -2$, i.e. since gravitation is based upon a second-order stress-energy tensor and consistent with spin-2 bosons that are massless and indistinguishable. This produces the following amplitude:
\begin{multline}
    A^{GR}_{++-} = \frac{[12]^{6}}{[23]^2 [31]^2} \delta^4 (\lambda^{(1)}_\alpha \tilde{\lambda}^{(1)}_{\dot{\alpha}} + \lambda^{(2)}_\alpha \tilde{\lambda}^{(2)}_{\dot{\alpha}} + \lambda^{(3)}_\alpha \tilde{\lambda}^{(3)}_{\dot{\alpha}}) \\
    = \frac{[12]^{6}}{[23]^2 [31]^2} \int d^4X^{\alpha \dot{\alpha}} e^{iX^{\alpha \dot{\alpha}} (\sum\limits_{i} \lambda^{(i)}_\alpha \tilde{\lambda}^{(i)}_{\dot{\alpha}})}
\end{multline}
Tracing the earlier procedure to Fourier transform ${\lambda}^{(1)}_{\alpha}$, ${\lambda}^{(2)}_{\alpha}$, and $\tilde{\lambda}^{(3)}_{\dot{\alpha}}$ into twistor components yields
\begin{multline}
    \hat{A}^{GR}_{++-} = [12]^6 \int d^4X^{\alpha \dot{\alpha}} d^2{\lambda}^{(1)}_{\alpha} d^2{\lambda}^{(2)}_{\alpha} da_1 da_2
  \{ 
    \\ \frac{\text{sgn}([12])[12]}{[23]^2[31]^2}e^{i\tilde{\mu}^{\alpha (1)} {\lambda}^{(1)}_{\alpha}} e^{i\tilde{\mu}^{\alpha (2)} {\lambda}^{(2)}_{\alpha}}
    e^{i{\mu}^{\dot{\alpha}(3)} (a_1 \tilde{\lambda}^{(1)}_{\dot{\alpha}} + a_2 \tilde{\lambda}^{(2)}_{\dot{\alpha}})} \\
    e^{iX^{\alpha \dot{\alpha}} (\lambda^{(1)}_\alpha \tilde{\lambda}^{(1)}_{\dot{\alpha}} + \lambda^{(2)}_\alpha \tilde{\lambda}^{(2)}_{\dot{\alpha}})}
    e^{iX^{\alpha \dot{\alpha}}\lambda^{(3)}_\alpha (a_1 \tilde{\lambda}^{(1)}_{\dot{\alpha}} + a_2 \tilde{\lambda}^{(2)}_{\dot{\alpha}})}
    \}
\end{multline}
\vspace{-0.8cm}
\begin{multline}
    \implies \hat{A}^{GR}_{++-} = \text{sgn}([12])[12]^5 \int d^4X^{\alpha \dot{\alpha}} \frac{da_1 da_2}{a^2_1a^2_2} \{ \\ \delta^2(\tilde{\mu}^{\alpha (1)} + X^{\alpha \dot{\alpha}} \tilde{\lambda}^{(1)}_{\dot{\alpha}}) e^{ia_1({\mu^{\dot{\alpha}}(3)}\tilde{\lambda}^{(1)}_{\dot{\alpha}} - {\tilde{\mu}^{\alpha}(1)} \lambda^{(3)}_\alpha )} \\
    \delta^2(\tilde{\mu}^{\alpha (2)} + X^{\alpha \dot{\alpha}} \tilde{\lambda}^{(2)}_{\dot{\alpha}})
    e^{ia_2({{\mu}^{\dot{\alpha}(3)}\tilde{\lambda}^{(2)}_{\dot{\alpha}}} - {\tilde{\mu}^{\alpha}(2)} \lambda^{(3)}_\alpha)}
    \}
\end{multline}
and using PV$\int^{\infty}_{-\infty} {da} \frac{e^{iax}}{(ca - b)^2} = \frac{-\pi e^{ibx/c}}{c^2}\lvert x \rvert$
results in the following non-perturbative 3-point amplitude \cite{Arkani_Hamed_2010}:
\begin{equation}
\label{a1-gr-3}
    \hat{A}^{GR}_{++-} = \lvert [12] \rvert 
    \lvert W^{(1)} \cdot Z^{(3)} \rvert
    \lvert W^{(2)} \cdot Z^{(3)} \rvert
\end{equation}
Similarly, the steps can all be repeated for the opposite helicity case (i.e. $h_1 = -2$, $h_2 = -2$, and $h_3 = +2$):
\begin{equation}
\label{a2-gr-3}
    \hat{A}^{GR}_{--+} = \lvert \langle 1 2 \rangle \rvert 
    \lvert W^{(3)} \cdot Z^{(1)} \rvert
    \lvert W^{(3)} \cdot Z^{(2)} \rvert
\end{equation}
3-point graviton scattering amplitudes have a simple yet significant difference when compared to gluons due to the absence of signum functions. Gravity permits a continuously varying probability space absolutely connected with interacting momenta rather than intrinsically binary +1 or --1 real amplitude found in pure Yang-Mills. 
\section{\label{sec:level3} Four-point scattering \\ in momentum and twistor space}
\noindent While 3-point interactions are an important starting point, these subamplitudes vanish for real momenta due to the absence of non-trivial kinematic invariants \cite{Bianchi_2008}, i.e. both $\langle i j \rangle = 0$ and $[ i j ] = 0$ are required since $\tilde{\lambda}_{\dot{\alpha}} = \pm (\lambda^*)_{\dot{\alpha}}$ to ensure $p_{\alpha \dot{\alpha}}$ is Hermitian \cite{cheung2017tasi}. Therefore, 4-point interactions generate the simplest non-vanishing scattering amplitudes, for example, in Minkowski space. Nevertheless, these subamplitudes are helpful building blocks because the product of two 3-point on-shell graphs exchanging zero helicity can be visualized as the scattering of 4 particles. Figure 2 displays a 4-point interaction in the s-channel, i.e. $s = (p^{(1)} + p^{(2)})^2 = \langle 1 2 \rangle [ 1 2 ] = 0$. The other possible tree-level factorization channels are given by the Mandelstam variables $u = (p^{(2)} + p^{(3)})^2 = \langle 2 3 \rangle [ 2 3 ]$ and $t = (p^{(1)} + p^{(3)})^2 = \langle 1 3 \rangle [ 1 3 ]$. (This similarly implies $s + t + u = \sum\limits_i m^2_i = 0$.) The s-channel residue, $R_s$, is \cite{cheung2017tasi}
\begin{equation}
    R_s = \lim\limits_{s \to 0} s A_{1234} = A_L A_R = A_{12I^+}A_{I^-34} 
\end{equation}
Taking all lines to be outgoing forces opposite helicity for $I$ on either side and, if 
\begin{equation}
\sum\limits_{i} h_L^{(i)} > 0 \text{ and } \sum\limits_{i} h_R^{(i)} < 0,
\end{equation}
then the scattering amplitude's residue becomes
\begin{multline} \label{Rs_tot}
    R_s = [12]^{h_1 + h_2 - h_{I^+}} [2I]^{h_2 + h_{I^+} - h_1} [I1]^{h_{I^+} + h_1 - h_2} \times \\ \langle I 3 \rangle^{h_4 - h_{I^-} - h_3} \langle 3 4 \rangle^{h_{I^-} - h_3 - h_4} \langle 4 I \rangle^{h_3 - h_4 - h_{I^-}}
\end{multline}
The following is prescribed without loss of generality
\begin{equation}
    \lambda^{(I)}_\alpha =  \lambda^{(1)}_\alpha \text{ and }
    \lambda^{(2)}_\alpha = c\lambda^{(1)}_\alpha
\end{equation}
\vspace{-0.4cm}
\begin{multline}  
    \sum\limits_{i} p^{(i)} = \lambda^{(1)}_\alpha (\tilde{\lambda}^{(1)}_{\dot{\alpha}} + c\tilde{\lambda}^{(2)}_{\dot{\alpha}} + \tilde{\lambda}^{(I)}_{\dot{\alpha}}) =  0 \\
    \implies \tilde{\lambda}^{(I)}_{\dot{\alpha}} = -\tilde{\lambda}^{(1)}_{\dot{\alpha}} - c\tilde{\lambda}^{(2)}_{\dot{\alpha}}
\end{multline}
where $c$ is a scalar (e.g. representing a little group transformation). Plugging these relations into (\ref{Rs_tot}), given helicity states $h_I = h_{I^+} = -h_{I^-}$, yields
\begin{figure}
\includegraphics{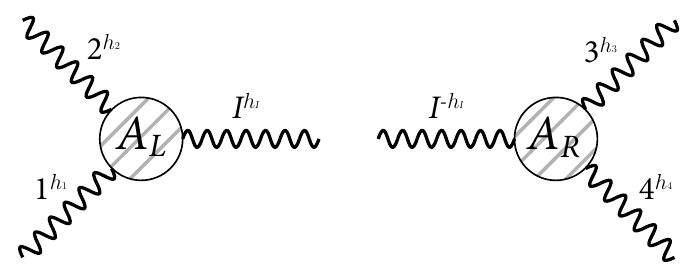}% Here is how to import EPS art
\caption{\label{fig:feyn_diag4} Example of four massless particles interacting via the s-channel, i.e. $\langle 1 2 \rangle [ 1 2 ] = 0$.}
\end{figure}
\begin{equation}
    \hspace{-0.15cm} R_s = \frac{[12]^{h_1 + h_2 + h_I - 1} \langle 1 3 \rangle^{-h_1 + h_2 - h_3 + h_4} }{\langle 1 2 \rangle \langle 2 3 \rangle^{-h_I - h_1 + h_2} \langle 3 4 \rangle^{h_I + h_3 + h_4} \langle 4 1 \rangle^{h_4 - h_3 - h_I}}
\end{equation}
\subsection{Gluon scattering}
\noindent With 4-point scattering amplitudes now defined, suppose gluons with $h_1 = h_3 = -1$ and $h_2 = h_4 = h_I = +1$ are interacting. The s-channel residue expectedly reduces to
\begin{equation} \label{RYM4}
    R^{YM}_{s-+-+} = \frac{s \langle 1 3 \rangle^4 }{\langle 1 2 \rangle \langle 2 3 \rangle \langle 3 4 \rangle \langle 4 1 \rangle}
\end{equation}
which is exactly the maximally helicity violating (MHV) amplitude given by the Parke-Taylor formula \cite{parke_taylor_physrev}. Note
\begin{equation}
    \frac{\langle 1 3 \rangle^4 }{\langle 1 2 \rangle \langle 2 3 \rangle \langle 3 4 \rangle \langle 4 1 \rangle} = \frac {\langle 1 3 \rangle^2 [ 2 4 ]^2}{s u},
\end{equation}
and summation over all factorization channels \cite{Heslop_2016} (i.e. not just $\langle 1 2 \rangle [ 1 2 ]$) outputs the full tree-level amplitude:
\begin{equation}
    \hspace{-0.4cm} A^{YM}_{-+-+} = {\langle 1 3 \rangle^2 [ 2 4 ]^2}({\frac{c_{st}}{st} + \frac{c_{tu}}{tu} + \frac{c_{us}}{us}}) %\delta^4 (\sum\limits_{i=1}^4\lambda^{(i)}_\alpha \tilde{\lambda}^{(i)}_{\dot{\alpha}})
\end{equation}
where the exact form of these constants can be found in \cite{cheung2017tasi} and satisfy the Jacobi identity as a naturally arising consistency condition due to factorization. The tree-level MHV amplitude will now be analyzed by Fourier transforming select spinor variables to reveal binarity extant in the transformed momentum space. To begin, let
\begin{equation} \label{trans2tilde}
    % \tilde{\lambda}^{(i)}_{\dot{\alpha}}
    \lambda^{(2)}_{\alpha} = b_1 \lambda^{(1)}_{\alpha} + b_3 \lambda^{(3)}_{\alpha}
\end{equation}
\begin{equation} 
\label{trans4tilde}
    \lambda^{(4)}_{\alpha} = d_1 \lambda^{(1)}_{\alpha} + d_3 \lambda^{(3)}_{\alpha}
\end{equation}
Fourier transforming with respect to $\tilde{\lambda}^{(1)}_{\dot{\alpha}}$, $\lambda^{(2)}_{\alpha}$, $\tilde{\lambda}^{(3)}_{\dot{\alpha}}$, and $\lambda^{(4)}_{\alpha}$ results in the given perturbative 4-point amplitude
\begin{multline}
    \hspace{-0.35cm} \hat{A}^{YM}_{-+-+} = \int \frac{d^2{\lambda}^{(2)}_{\alpha} d^2{\lambda}^{(4)}_{\alpha} \langle 1 3 \rangle^4 }{\langle 1 2 \rangle \langle 2 3 \rangle \langle 3 4 \rangle \langle 4 1 \rangle} d^4X^{\alpha \dot{\alpha}}
    d^2 \tilde{\lambda}^{(1)}_{\dot{\alpha}} 
    d^2 \tilde{\lambda}^{(3)}_{\dot{\alpha}} \{ \\
    e^{i{\mu}^{\dot{\alpha}(1)} \tilde{\lambda}^{(1)}_{\dot{\alpha}}} e^{i\tilde{\mu}^{\alpha (2)} {\lambda}^{(2)}_{\alpha}} e^{i{\mu}^{\dot{\alpha}(3)} \tilde{\lambda}^{(3)}_{\dot{\alpha}}} e^{i\tilde{\mu}^{\alpha (4)} {\lambda}^{(4)}_{\alpha}} 
    e^{iX^{\alpha \dot{\alpha}}
    (\sum\limits_{i} \lambda^{(i)}_\alpha \tilde{\lambda}^{(i)}_{\dot{\alpha}})} \}
\end{multline}
% \vspace{-0.85cm}
% \begin{equation*}
%     \displaystyle\left\Updownarrow\vphantom{\int}\right.
% \end{equation*}
% \vspace{-0.55cm}
% \begin{multline*}
%     \hspace{-0.35cm} \hat{A}^{YM}_{-+-+} = \int d^4X^{\alpha \dot{\alpha}} \frac{db_1 db_3 dd_1 dd_3}{b_1 b_3 d_1 d_3} d^2 \tilde{\lambda}^{(1)}_{\dot{\alpha}}
%     d^2 \tilde{\lambda}^{(3)}_{\dot{\alpha}} \{ \langle 1 3 \rangle^2 e^{i{\mu}^{\dot{\alpha}}^{(1)} \tilde{\lambda}^{(1)}_{\dot{\alpha}}} \\ e^{i\tilde{\mu}^{\alpha}^{(2)} (b_1 \lambda^{(1)}_{\alpha} + b_3 \lambda^{(3)}_{\alpha})} e^{i{\mu}^{\dot{\alpha}}^{(3)} \tilde{\lambda}^{(3)}_{\dot{\alpha}}} e^{i\tilde{\mu}^{\alpha}^{(4)} (d_1 \lambda^{(1)}_{\alpha} + d_3 \lambda^{(3)}_{\alpha})}
%     e^{iX^{\alpha \dot{\alpha}} \lambda^{(1)}_\alpha \tilde{\lambda}^{(1)}_{\dot{\alpha}}} \\
%     e^{iX^{\alpha \dot{\alpha}} (b_1 \lambda^{(1)}_{\alpha} + b_3 \lambda^{(3)}_{\alpha}) \tilde{\lambda}^{(2)}_{\dot{\alpha}}}
%     e^{iX^{\alpha \dot{\alpha}} \lambda^{(3)}_\alpha \tilde{\lambda}^{(3)}_{\dot{\alpha}}}
%     e^{iX^{\alpha \dot{\alpha}} (d_1 \lambda^{(1)}_{\alpha} + d_3 \lambda^{(3)}_{\alpha}) \tilde{\lambda}^{(4)}_{\dot{\alpha}}}
%     \}
% \end{multline*} 
Applying the following expression 
\begin{multline*}
\hspace{-0.4cm}
    \int d^4X^{\alpha \dot{\alpha}}  \delta^2({\mu}^{\dot{\alpha}(i)} + X^{\alpha \dot{\alpha}} {\lambda}^{(i)}_{\alpha})
    \delta^2({\mu}^{\dot{\alpha}(j)} + X^{\alpha \dot{\alpha}} {\lambda}^{(j)}_{\alpha}) \rightarrow \frac{1}{\langle i j \rangle^2}
\end{multline*}
with the previously enumerated results produces \cite{Arkani_Hamed_2010}
\begin{multline}
\label{a-YM4-final}
    \hat{A}^{YM}_{-+-+} = \text{sgn}(Z^{(1)} \cdot W^{(2)}) \text{sgn}(W^{(2)} \cdot Z^{(3)}) \\ \text{sgn}(Z^{(3)} \cdot W^{(4)}) \text{sgn}(Z^{(1)} \cdot W^{(4)})
\end{multline}
This gluon scattering interaction has non-vanishing amplitude even for real $p_{\alpha \dot{\alpha}}$ and reveals its inherently binary nature through its composition of signum functions which effectively constrains outputs to be either +1 or --1 \cite{Arkani_Hamed_2010}.
\subsection{Graviton scattering}
\noindent For graviton scattering, such as $h_1 = h_3 = -2$ and $h_2 = h_I = h_4 = +2$, the s-channel residue is
\begin{equation}
    R^{GR}_{s-+-+} = \frac{s^2  \langle 1 3 \rangle^8 }{\langle 1 2 \rangle^2 \langle 2 3 \rangle^2 \langle 3 4 \rangle^2 \langle 4 1 \rangle^2}
\end{equation} 
and this can be converted into the corresponding amplitude (since $u = -t$ when $s = 0$)
\begin{equation}
    A^{GR}_{-+-+} = -\frac {\langle 1 3 \rangle^4 [ 2 4 ]^4}{stu} \delta^4 (\sum\limits_{i=1}^4\lambda^{(i)}_\alpha \tilde{\lambda}^{(i)}_{\dot{\alpha}})
\end{equation}
which is in fact the complete 4-point tree-level graviton scattering amplitude as it contains all factorization channels. This is the starting point for now shifting select momenta into twistor variables ($\mu_{\dot{\alpha}}$, $\tilde{\mu}_\alpha$). Beginning with
\begin{equation}
    A^{GR}_{-+-+} = \frac {\langle 1 2 \rangle [ 1 2 ] \langle 1 3 \rangle^8}{\langle 1 2 \rangle^2 \langle 2 3 \rangle^2 \langle 3 4 \rangle^2 \langle 4 1 \rangle^2} \delta^4 (\sum\limits_{i=1}^4\lambda^{(i)}_\alpha \tilde{\lambda}^{(i)}_{\dot{\alpha}})
\end{equation}
and using the same transformations (\ref{trans2tilde}) and (\ref{trans4tilde}) with
\begin{equation}
    \langle 1 3 \rangle [ 1 2 ] = - \langle 3 1 \rangle [1 2] = \langle 3 \lvert p_2 + p_3 + p_4 \rvert 2 ]
    = \langle 3 4 \rangle [ 4 2 ]
\end{equation} 
\begin{multline}   
\label{gr_scat}
\hspace{-0.45cm} \implies \hat{A}^{GR}_{-+-+} = \langle 1 3 \rangle [ 2 4 ] \int d^4X^{\alpha \dot{\alpha}} \frac{db_1 db_3 dd_1 dd_3}{b^2_1 b_3 d_1 d^2_3} d^2 \tilde{\lambda}^{(1)}_{\dot{\alpha}} \\
    d^2 \tilde{\lambda}^{(3)}_{\dot{\alpha}} \{ e^{i{\mu}^{\dot{\alpha}(1)} \tilde{\lambda}^{(1)}_{\dot{\alpha}}} e^{i\tilde{\mu}^{\alpha (2)} (b_1 \lambda^{(1)}_{\alpha} + b_3 \lambda^{(3)}_{\alpha})}  e^{i\tilde{\mu}^{\alpha (4)} (d_1 \lambda^{(1)}_{\alpha} + d_3 \lambda^{(3)}_{\alpha})} \\ \hspace{-0.15cm}
    e^{i{\mu}^{\dot{\alpha}(3)} \tilde{\lambda}^{(3)}_{\dot{\alpha}}}
    e^{iX^{\alpha \dot{\alpha}} \lambda^{(1)}_\alpha \tilde{\lambda}^{(1)}_{\dot{\alpha}}} 
    e^{iX^{\alpha \dot{\alpha}} (b_1 \lambda^{(1)}_{\alpha} + b_3 \lambda^{(3)}_{\alpha}) \tilde{\lambda}^{(2)}_{\dot{\alpha}}} \\
    e^{iX^{\alpha \dot{\alpha}} \lambda^{(3)}_\alpha \tilde{\lambda}^{(3)}_{\dot{\alpha}}}
    e^{iX^{\alpha \dot{\alpha}} (d_1 \lambda^{(1)}_{\alpha} + d_3 \lambda^{(3)}_{\alpha}) \tilde{\lambda}^{(4)}_{\dot{\alpha}}}
    \}
\end{multline} 
After evaluating the integral in (\ref{gr_scat}), the full tree-level amplitude for interacting gravitons is consequently
\begin{multline}
\label{a-GR4-final}
    \hat{A}^{GR}_{-+-+} = \langle 1 3 \rangle [ 2 4 ] | Z^{(1)} \cdot W^{(2)} | | Z^{(3)} \cdot W^{(4)}| \\ \text{sgn}(W^{(2)} \cdot Z^{(3)}) \text{sgn}(Z^{(1)} \cdot W^{(4)})
\end{multline}
This 4-point interaction of gravitons remarkably reveals both manifestly binary and non-binary variables in its amplitude contrary to 3-point scattering given by (\ref{a1-gr-3}) and (\ref{a2-gr-3}). Furthermore, (\ref{a-GR4-final}) contains signum functions in addition to products of the kinematic variables while (\ref{a-YM4-final}) consists of only signum functions. This implies interacting gluons retain strict binarity in their 4-point scattering amplitude while the 4-point gravitational scattering amplitude has both binary and non-binary features. The latter amplitude is inevitably non-binary in general since a signum function multiplied by a non-binary function is not necessarily constrained to discrete values. Therefore, when taking the modulus squared of amplitudes to calculate scattering probabilities, the pure Yang-Mills scenario only exists in discrete pockets of probability space while essentially independent of momenta magnitudes. The gravitational interaction's probability amplitude on the other hand is not necessarily confined to discrete intervals and varies directly with interacting momenta.  
\vspace{-0.55cm}
\section{Conclusion} 
\vspace{-0.35cm}
\noindent The full 3-point scattering amplitudes for massless particles are reviewed for Yang-Mills and gravity from a common origin of on-shell diagrams. The 4-point tree-level amplitudes are then constructed and correspond to the first non-trivial scattering events, for example, in Minkowski spacetime where momenta are real. It is revealed that tree-level interactions with gluons contain only terms with distinct binarity associated with the +1 or --1 in the signum functions, while gravitational interactions involving only scattering of gravitons do not solely consist of signum functions, {\it but they are nevertheless present}. This denotes an intrinsic difference in the probability space of observable interactions due to the momentum structure of the interacting particles. Namely, it is evident through Fourier transforming spinor variables in momentum space that pure gluon (YM) and graviton (GR) scattering events occupy categorically discrete and continuous regions in probability space, respectively. 
\vspace{-0.3cm}
\begin{acknowledgments}
\vspace{-0.2cm}
\noindent I wish to thank N. Arkani-Hamed for his lectures taught at Harvard University and insightful discussions throughout these months with S. Alipour-fard. Supported by the Natural Sciences and Engineering Research Council of Canada (NSERC) doctoral postgraduate scholarship.
\end{acknowledgments}

\nocite{*}

\bibliography{apssamp}% Produces the bibliography via BibTeX.
\end{document}